\documentclass[12pt]{article}
\begin{document}
\title{Comment on Hervik's comment (astro-ph/0504071) on evidence of vorticity  
and shear of the Universe (astro-ph/0503213)}
\author{D. PALLE \\
Zavod za teorijsku fiziku \\
Institut Rugjer Bo\v skovi\'c \\
Po\v st. Pret. 180, HR-10002 Zagreb, CROATIA}
\date{ }
\maketitle

\vspace{5 mm}
{\it
Hervik's comments on my criticism of the paper of Jaffe et al 
are irrelevant or wrong.
}
\vspace{5 mm}

Jaffe et al \cite{jaffe} performed recently the fit of 
WMAP data based on the work of Barrow et al \cite{barrow}.
They claim \cite{jaffe} the existence of the nonvanishing vorticity and
shear of the Universe.
The spacetime vorticity \cite{goedel} in the model \cite{barrow}
vanishes by definition, and this was the statement in my
comment \cite{palle}.
The spacetime vorticity vanishes even after the introduction 
of the angle of tilt that induces fluid vorticity \cite{king}.
However, the model discussed in \cite{barrow} contains
vanishing angle of tilt, thus also vanishing fluid vorticity.
Barrow et al \cite{barrow} define their own "vorticity", with
unclear physical notion and relevance.

The correct treatment of perturbations in General 
Relativistic cosmology beyond FRW models is introduced and
discussed by Ellis and Bruni \cite{ellis}, few
years after paper of Barrow et al had been published.
One cannot work properly in the model beyond FRW, and then assume
during some derivations "FRW background" \cite{hervik}.
The gauge noninvariant formulae are the source of 
the embarrassing results of Jaffe et al \cite{jaffe}: extremely small
shear beyond one-year WMAP data sensitivity.

SAPIENTI SAT.

\end{document}